\documentclass[notoc]{JHEP3} 

\newcommand{\ie}{{\it i.e.}}
\newcommand{\eg}{{\it e.g.}}
\newcommand{\Tr}{{\rm  Tr }}
\newcommand{\be}{\begin{equation}}
\newcommand{\ee}{\end{equation}}
\newcommand{\ba}{\begin{eqnarray}}
\newcommand{\ea}{\end{eqnarray}}

\title{Trace formulas for Annuli}

\author{A. N. Schellekens \\
{\it NIKHEF, P.O. Box 41882, 1009$\,$DB Amsterdam,
The Netherlands } }
\author{
Ya. S. Stanev \thanks{On Leave from Institute for Nuclear Research and Nuclear  
Energy, Bulgarian
Academy of Sciences, BG-1784 Sofia, BULGARIA.}~\\
{\it  Dipartimento di Fisica
Universit{\`a} di Roma \ ``Tor Vergata'' \\
I.N.F.N.\ - \ Sezione di Roma \ ``Tor Vergata'' \\
Via della Ricerca Scientifica 1 ,\ \
00133 \ Roma \ \ ITALY } }

\received{\today}       
\revised{\today}
\accepted{\today}       

\preprint{\hepth{0108035}}  

\abstract{Assuming the completeness condition for boundaries we
derive trace formulas for the annulus coefficients in 2-dimensional conformal field
theory. We also derive polynomial equations that relate the annulus, Moebius
and Klein bottle coefficients, and conjecture an annulus trace formula that is 
sensitive to the orientation of the boundaries.}


\begin{document}

\section{Introduction}

\def\Zbf{{\bf Z}}

To compute perturbative open string spectra using
(unitary, rational) conformal field theory
one needs to know the following data~\cite{Cardy},~\cite{Rome}: a multiplicity  
matrix $Z_{ij}$
that defines the torus partitition function, a set $\{a\}$
of allowed boundary conditions, a set of coefficients
$B_{ma}$ that describe the reflection of bulk fields at boundary $a$,
and a set of coefficients $\Gamma_m$ that decribe the behavior of bulk fields
in the presence of a crosscap. Given these data one can compute all closed
and open string partition functions. Of course,
much more information is needed to
compute correlation functions.

Three different labels where introduced here: The labels $i,j,\ldots$ refer
to primary fields of the bulk CFT; the labels $a,b,\ldots$ indicate distinct
boundary conditions, and the labels $m,n,\ldots$ correspond to those
bulk fields that can appear in the transverse channel coupling
to boundaries and/or crosscaps. More precisely the latter correspond
to Ishibashi boundary and crosscap states that preserve the chiral algebra.
The complete set of such states corresponds to the bulk fields that are
paired with their charge conjugate, \ie\ those for which  $\Zbf_{ii^c} \not=0$.
In this paper we will consider arbitrary symmetric modular invariant
matrices $Z_{ij}$. In particular this includes matrices with a non-trivial
kernel, implying an extension of the chiral algebra. The Ishibashi states
we will use are only required to preserve the original, unextended algebra,
so that the boundaries and crosscaps may break part of the extended symmetry  
~\cite{SB}.

In principle, one would like to determine all allowed boundary and crosscap
coefficients
given a modular invariant $Z_{ij}$. This problem can be transformed from a
problem over the real numbers to a problem in terms of bounded
integers by describing it in terms of annulus, Moebius and Klein bottle
coefficients. Just like the search for modular invariants this is still a
difficult problem to solve in general, and indeed the goal of this
paper is more modest. We merely want to formulate a set of polynomial
equations and trace formulas that the solutions should
satisfy. Part
 of these results can be derived under mild assumptions from the
``completeness condition" to be discussed below, others are conjectures
which we can only prove in special cases. The complete solution given in
~\cite{FOE}
for all Klein bottle choices and all simple current modular invariants serves
as a useful guiding principle for the general case, as well as a
non-trivial test
for the conjectures.

A complication that is usually associated with extensions of the
chiral algebra is the appearance of multiplicities larger than one. If
a matrix element $Z_{ii^c}$ is larger than one, the corresponding Ishibashi
states are degenerate and we must introduce an additional degeneracy label
$0 \leq \alpha(i) < Z_{ii^c}$ (to simplify the notation we shall omit in the rest 
of the paper the dependence of $\alpha$ on $i$).
In the standard computation of the annulus, Moebius
and Klein bottle amplitudes we cannot rely on purely representation-theoretic
arguments anymore, because this gives no information regarding the overlap
of states within the same degeneracy space. A general parametrization yields
the following expressions\footnote{The degeneracy matrices first appeared in  
~\cite{FOE}.
As we show here, they can be transformed to the identity, but in that basis
the reflection and crosscap coefficients presented in ~\cite{FOE} become
more complicated.}
\ba
 A^i_{ab} &=& \sum_{m,\alpha,\beta} S^i_{~m} g_m^{\alpha\beta}
B_{(m,\alpha)a}B_{(m,\beta)b} \\
K^i &=& \sum_{m,\alpha,\beta} S^i_{~m} k_m^{\alpha\beta}
\Gamma_{(m,\alpha)}\Gamma_{(m,\beta)} \\
M^i_{~a} &=& \sum_{m,\alpha,\beta} P^i_{~m} h_m^{\alpha\beta}
B_{(m,\alpha)a}\Gamma_{(m,\beta)}
\ea
where $S$ is the standard modular matrix, while $P=\sqrt{T}ST^2S\sqrt{T}$  
~\cite{Pmat}.
Note that $k_m$ and $g_m$ are symmetric matrices. This means in particular
that they have a square root, which we can absorb in the definition
of the coefficients $B$ and $\Gamma$. In this way we can see that without loss
of generality $k$ and $g$ may be replaced by the identity matrix. This changes
$h$ to a new matrix $h'$.
Having done that,
we may allow, in terms of the new coefficients,
orthogonal rotations in degeneracy space,
which do not alter the Klein bottle and the annulus:
$$ B_{(m,\alpha)a} \to \sum_{\beta} W^m_{\alpha\beta} B_{(m,\beta)a} \ \ \ \ ,\ \ \
 \Gamma_{(m,\alpha)} \to \sum_{\beta} V^m_{\alpha\beta} \Gamma_{(m,\beta)} $$
This changes the matrix $h'$ to
 $$ h_m^{\prime\prime}=(W^m)^T h_m^{\prime} V^m \ .$$
This allows us to diagonalize $h_m''$. The eigenvalues will be denoted
$\lambda^m_{\alpha}$. It is now instructive to transform to the transverse
channel. Then the amplitudes are
\ba
\tilde A_{ab} &=& \sum_{m,\alpha}
B_{(m,\alpha)a}B_{(m,\alpha)b}{\cal X}_m \\
\tilde K &=& \sum_{m,\alpha}
\Gamma_{(m,\alpha)}\Gamma_{(m,\alpha)}{\cal X}_m \\
\tilde M_{~a} &=& \sum_{m,\alpha}  \lambda_m^{\alpha}
B_{(m,\alpha)a}\Gamma_{(m,\alpha)}\hat{\cal X}_m\ , \label{eq:Transverse}
\ea
where ${\cal X}_m$ is a character (with the usual arguments and the
usual definition of $\hat{\cal X}$).
In the absence of degeneracies the Moebius amplitude is the ``geometric
mean" of the annulus and the Klein bottle.
The result (\ref{eq:Transverse}) violates
this geometric mean principle  unless $\lambda_m^{\alpha}$
 is just a sign. But then we can absorb it into the
definition of the crosscap coefficients without changing the Klein bottle. Hence
if we adopt the geometric mean principle we may from now on assume that
all $\lambda$'s are equal to 1.

\section{Completeness}

In string theory the identity character
(which in our notation corresponds to the label $0$)
gives rise to gauge bosons. The resulting
gauge groups are only identifiable with $SO(N)$, $Sp(N)$ or $U(N)$ for generic $N$
if $A^0_{~ab}$ is an involution. This implies
\be
 \sum_{i,\alpha}S_{i0} [B_{(i,\alpha)a} B_{(i,\alpha)b} ] = \delta_{ba^c}\ ,
\ee
where $a^c$ is, by definition, the boundary conjugate to $a$. The fact
that boundary conjugation must be an involution implies that the
reflection coefficients
$R_{(m,\alpha),a}=\sqrt{S_{0m}} B_{(m,\alpha),a} $ are orthogonal
$$ \sum_{(m,\alpha)} R_{(m,\alpha),a}R_{(m,\alpha),b^c} = \delta_{ab} $$
This puts an upper limit on the number of distinct boundaries that
can appear in a given theory: the number cannot exceed the number
of Ishibashi states, counted according to their degeneracy $\Zbf_{ii^c}$.

Just like modular invariance is a completeness condition for operators
in the bulk CFT, it is natural to postulate a completeness condition
for boundaries in the open string case, namely that
the
upper bound is saturated. This implies that $R$ is a square matrix,
and that there must exist an inverse $\hat R$ so that
$$ \sum_a \hat R_{(m,\alpha),a}R_{(n,\beta),a}=
\delta_{nm}\delta_{\alpha\beta} $$
Multiplying by $R_{(n,\beta),b}$ and summing over $(n,\beta)$ we find
then that $\hat R_{(m,\alpha),a} = R_{(m,\alpha),a^c}$, so that
\be
 \sum_a R_{(m,\alpha)}^{~~~~~~a}R_{(n,\beta),a}=
\delta_{nm}\delta_{\alpha\beta} \label{eq:cmplt}
\ee
Here we use raised indices to indicate boundary conjugation.
{}From this form of the completeness condition one can
straightforwardly derive another well-known expression
\be \sum_b A_{ia}^{~~b} A_{jb}^{~~c} = \sum_k N_{ij}^{~~k}A_{ka}^{~~c} \  
,\label{eq:AANA}
\ee
where $N_{ij}^{~~k}$ are the fusion coefficients, expressed in terms
of the Verlinde formula. This formula has a heuristic interpretation
in terms of two ways of counting the number of couplings of the
correlator $\langle a \mid \Phi^i \Phi^j \mid c \rangle $, on the one hand via  
fusion, and
on the other hand via insertion of a complete set of boundary states.
It is hard to turn this heuristic argument into  a rigorous proof, but
we will not need
this interpretation anyway.

The completeness condition
 (\ref{eq:AANA}) was first written down in ~\cite{CMPLT} \footnote{The same
formula appeared earlier in ~\cite{DiFZ}, but in a different context.} and has
been the starting point of a lot of later work (see \eg\  
~\cite{Zuber},~\cite{Gann}). It
should be emphasized that completeness for boundaries is not on equal
footing with completeness for bulk operators, \ie modular invariance, as
a consistency condition.
Whereas a violation of the latter leads to clearly identifiable
inconsistencies in string theory,
there are, generically,  no obvious inconsistencies associated with violating  
completeness
for boundaries. Indeed, in string theory boundaries are counted with
Chan-Paton multiplicities, and no principle is violated if some of
these multiplicities vanish (as is often required by tadpole cancellation).
On the other hand it is known from many examples that completeness of boundaries
corresponds correctly to completeness of the set of branes, as can be verified
through dualities. Furthermore in conformal field theory
complete sets of boundaries have been found in many cases
including the large class of simple current modular invariants ~\cite{FOE}.
In all well-studied cases
completeness emerges as a statement regarding the
complete set of one-dimensional
representations of a commutative algebra (the fusion algebra in the
``Cardy case" ~\cite{Cardy}, more general classifying algebras~\cite{ClassAlg}  
in other cases).
Presumably the correct mathematical setting to deal with the general
case is still missing, but on the basis of current experience it seems
reasonable to assume completeness as a consistency condition for boundary CFT.

Formula (\ref{eq:AANA}) is not only a consequence of completeness, but, under
a very mild assumption, equivalent to it.
First of all we start with the definition of the annulus coefficients
and derive from it
\be \sum_j A^j_{~ab}S_{jm} = \sum_{\alpha} {R_{(m,\alpha)a} R_{(m,\alpha)b}  \over
 S_{0m} } \label{eq:baseq}
\ee
We multiply both sides with $A^{\ell}_{~b^cc}$ and sum over $b$:
\be \sum_{j,b} A^j_{~ab}A^{\ell}_{~b^cc}S_{jm} = \sum_{\alpha,b}
{R_{(m,\alpha)a} R_{(m,\alpha)b}  \over
 S_{0i} }A^{\ell}_{~b^cc}
 \ee
Consider the left hand side. Using (\ref{eq:AANA})
 and the Verlinde formula and finally once again (\ref{eq:baseq}) we get
$$ \sum_{j,b} A^j_{~ab}A^{\ell}_{~b^cc}S_{jm}
=    { S^{\ell}_m \over S_{0m} }
\sum_{\alpha} {R_{(m,\alpha)a} R_{(m,\alpha)c}  \over
 S_{0m} }\ . $$

Combining this with the right hand side we obtain
$$
\sum_{\alpha} R_{(m,\alpha)a} \left[  { S^{\ell}_m \over S_{0m} }
 R_{(m,\alpha)c}  - \sum_{b} R_{(m,\alpha)b}
A^{\ell}_{~b^cc}  \right] $$
Note that there is no summation on $m$ here! For fixed $\ell, m$ and $c$ we find
here a set of conditions of the form
$$ \sum_{\alpha} X_{\alpha}(m,\ell,c) V^a_{\alpha}(m)= 0 \ ,$$
where $V$ stands for $R$. We have such a condition for each $a$.
The condition says that the vector $X_{\alpha}$ must be orthogonal to
the set of vectors $V^a_{\alpha}$, where $a$ runs over the set of boundaries.
If the
vectors $V^a_{\alpha}$, considering all $a$, span the degeneracy space of $m$,
then this set equations implies that $X=0$.  If on the other hand the equations
do not imply $X=0$ (for some $m,\ell$ and $c$), then there must be at least
one direction in the degeneracy space of $m$ that is orthogonal to all
$V^a_{\alpha}$. We can then make a rotation in the degeneracy space of $m$
so that this orthogonal direction coincides, for example, with $\alpha=0$.
Then $V^a_0 = 0$ for all $a$, or in other words $R_{(m,0)a}=0$ for all $a$,
so that one Ishibashi label does not couple to any boundary.

Conversely, if we assume that all Ishibashi labels couple to at least one
boundary, we find that $X=0$ (Here ``all Ishibashi labels" means ``there is
no basis in the degeneracy space such that one Ishibashi label completely  
decouples").

The condition $X=0$ reads
$${ S^{\ell}_m \over S_{0m} }
 R_{(m\alpha)c}  - \sum_{b} R_{(m,\alpha)b}
A^{\ell}_{~b^cc}=0 $$
Now we multiply by $S_{\ell n}$ and sum over $\ell$:
\ba
 {R_{(n,\alpha)c}\over S_{0n} } \delta_{mn}
  &=& \sum_{b} R_{(m,\alpha)b}
\sum_{\ell} S_{\ell n}A^{\ell}_{~b^cc}  \nonumber \\
&=& \sum_{b} R_{(m,\alpha)b} \sum_{\beta}
{R_{(n,\beta)b^c}R_{(n,\beta)c} \over S_{0n}}
\ea
This can be written as
$$ \sum_{\beta} R_{(n,\beta)c} \left[\delta_{\alpha\beta} \delta_{mn}
-\sum_{b} R_{(m,\alpha)b}
R_{(n,\beta)b^c}\right]   $$
Exactly the same ``non-decoupling" assumption regarding Ishibashi labels
now yields (\ref{eq:cmplt}).

\section{Polynomial equations}

{}From (\ref{eq:cmplt}) one easily derives the following polynomial equations
for the one-loop open string amplitudes (here $Y_{ijk}=\sum_n
S_{in}P_{jn}P_{kn}/S_{0n}$ are integers ~\cite{Ytens})
\be \sum_b A_{ia}^{~~b}A_{jb}^{~~c}=
\sum_k N_{ij}^{~~k} A_{ka}^{~~c} \label{eq:AANA2} \ee
\be \sum_b A_{iab}M_{j}^{b}=\sum_l Y_{ij}^{~~l} M_{la} \label{eq:AMYM} \ee
\be \sum_a M_i^{a}M_{ja}= \sum_l Y^l_{~ij}K_l \label{eq:MMYK} \ee
There are two more equations that can only be derived if
there are no degeneracies in $Z_{mm^c}$. Unlike the previous
equations they involve only summations over bulk labels.
\be \sum_i A_{iab}A^{i}_{cd}=
\sum_i A_{iac} A^{i}_{bd} \label{eq:AAAA} \ee
\be  \sum_i M_{ia}M^{i}_{b}=\sum_i A_{iab}K^i \label{eq:MMAK} \ee
Although in particular the first of these has a nice duality-like
graphical interpretation, it does not hold in general, and is in fact
explicitly violated by some of the cases discussed in ~\cite{FOE}.

Equations (\ref{eq:AANA2}), (\ref{eq:AMYM}) and (\ref{eq:MMYK}) can be used in  
attempts to determine
boundary and crosscap coefficients in rational CFT's with exceptional
modular invariants and/or non-standard Klein bottle choices.

\section{Trace identity}

{}From the definition of the annulus and (\ref{eq:cmplt}) we derive immediately
\be \sum_b A_{ib}^{~~b} = \sum_{\ell,\alpha} {S_{i\ell}\over S_{0\ell}}
\delta_{\alpha\alpha}
\label{eq:TraceOne} \ee
The sum over $\alpha$ is equal to $Z_{\ell\ell^c}$.
This may be written as
$$  Z_{\ell\ell^c}=(ZC)_{\ell\ell}=(ZSS)_{\ell\ell}=(SZS)_{\ell\ell} $$
where in the last step modular invariance was used. The two factors $S$
combine with those in (\ref{eq:TraceOne}) to yield a fusion coefficient. The  
final result is
\be \sum_b A_{ib}^{~~b} = \sum_{j,k} N_i^{~jk} Z_{jk} \label{eq:TraceTwo} \ee

Note that (\ref{eq:TraceOne}) is an interesting and non-trivial test for the  
C-diagonal
part of
potential modular invariants. This is independent of the existence of a boundary
CFT, since in the form (\ref{eq:TraceTwo}) the right hand side is manifestly integer.

This trace identity can be extended to higher order using (\ref{eq:AANA}),
and in general one gets
$$ \Tr(A_{i_1}A_{i_2}\ldots A_{i_n})=\Tr(N_{i_1}N_{i_2}\ldots N_{i_n}Z) $$
where all traces and matrix multiplications are in terms of implicit
raised and lowered indices.

Another immediate consequence of the trace identity  (\ref{eq:TraceOne}) is that 
\be
{\cal N}^{(g)}_{{i_1}{i_2} \dots {i_n}} =
\sum_p { S_{i_1 p} \over S_{0 p}} { S_{i_2 p} \over S_{0 p}} \dots
{ S_{i_n p} \over S_{0 p}}  {1 \over (S_{0 p})^{2(g-1)}} Z_{p p^c} \quad
\label{eq:TraceYone}
\ee
are nonnegative integers for any integer $g \geq 1$.
Indeed, introducing the standard higher genus Verlinde coefficients
\be
{ N}^{(g)}_{{i_1}{i_2} \dots {i_n}} =
\sum_p { S_{i_1 p} \over S_{0 p}} { S_{i_2 p} \over S_{0 p}} \dots
{ S_{i_n p} \over S_{0 p}}  {1 \over (S_{0 p})^{2(g-1)}}  \quad
\label{eq:VerNgenus}
\ee
one easily shows the identity
$$
{\cal N}^{(g+1)}_{{i_1}{i_2} \dots {i_n}} = \sum_k
{N}^{(g)~~~~~k}_{{i_1}{i_2} \dots {i_n}} \ \sum_b A_{kb}^{~~b}
$$
which explains also the constraint $g \geq 1$.

\section{Orientation-sensitive trace identities}

All of the previous formulas concerned the annulus amplitudes
$A_{ia}^{~~b}$.
In the simple current case there are in general for each modular
invariant several choices of crosscap coefficients, each with their own
complete set of boundary coefficients. It turns out that the
amplitudes $A_{ia}^{~~b}$ are not sensitive to these differences,
essentially because the only effect of the orientation-dependent
choices is to change the boundary charge conjugation $A^0_{~ab}$, which
drops out in $A_{ia}^{~~b}$.
The quantities $A^i_{~ab}$, the physically relevant ones in string theory,
are however sensitive to these differences.

\def\mod{~{\rm mod}~}
Let us first compute
$$ \sum_a A^0_{~aa} $$
To compute this trace we make use of the open string partition function
integrality condition
$$ A^i_{aa} \geq | M^i_{~a} | \hbox{~and~} A^i_{aa}=M^i_{~a} \mod 2$$
For $i=0$ the only possibilities for $A^0_{~aa}$ are 0 or 1, and hence
$$ A^0_{~aa}=(M^0_a)^2 $$
We sum this using (\ref{eq:MMYK}) (note that $M^0_a=M^0_{a^c}$, because
both vanish if $a\not=a^c$). Then we get
$$ \sum_a A^0_{~aa}=\sum_{\ell} Y^{\ell}_{~00} K_{\ell} $$
This equation has a simple interpretation, in particular if we write it as
$$ {1 \over 2} (\sum_a(\delta_{aa}+ (M^0_a)^2) =
{1 \over 2} (\sum_{\ell} (Z_{\ell\ell^c} + Y^{\ell}_{~00} K_{\ell})) $$
The left-hand side is the number of CP gauge groups. The right-hand side is  
the number
of Ishibashi scalars that survive the Klein bottle projection.
To see the latter, note that $Y^{\ell}_{~00}$ is equal to the
Frobenius-Schur indicator of primary $i$ ~\cite{Kbtl},~\cite{Bantay}, which vanishes
for complex fields. Therefore
if $\ell\not=\ell^c$ $Y^{\ell}_{~00}$ vanishes, so that
these states contribute with a factor ${1 \over 2}$, precisely the reduction of their
multiplicity.
If the Klein bottle equals the
FS indicator and is non-zero, then $\ell=\ell^c$. Each such state contributes a
factor 1. If the Klein bottle has the opposite sign, the state is projected
with a sign opposite the FS-indicator, which implies that the singlet is
projected out. It was tacidly assumed here that the degeneracies are 0 or 1.
For higher multiplicities the interpretation is essentially the same.

We may write this identity also as
\be \sum_{\ell} \sum_{a , \alpha}
R_{(\ell,\alpha)a}R_{(\ell,\alpha)a}=\sum_{\ell} Y^{\ell}_{~00} K_{\ell}
\label{eq:traceAzero} \ee
Although we have derived this with a summation over $\ell$,
it turns out that in all cases studied so far
this relation holds also {\it without summation}!

This
conjecture can be rewritten in terms of the
 trace identity
\be \sum_a A^i_{aa}=\sum_{\ell} {S^{i\ell}\over S_{0\ell}} Y^{\ell}_{~00} K_{\ell} 
\label{eq:traceconj} \ee
Note that the right hand side is not manifestly integer. Therefore
this relation -- if true -- implies a powerful constraint on
possible Klein bottle choices.

Unfortunately we have been unable to prove this trace-formula in general,
but we can give additional support for it in special cases using
the classifying algebra, which follows from the sewing constraints.
The classifying algebra reads
\be R_{(m,\alpha)a}R_{(n,\beta)a}=\sum_{\ell,\gamma}
X_{(m,\alpha)(n,\beta);(\ell,\gamma)}
R_{(\ell,\gamma)a}R_{0a} \ ,
\label{eq:classif} \ee
where $X_{pq;r}$ are the structure constants, which are symmetric in
$p$ and $q$.
Note that $0$ does not have a degeneracy. If we make the
very plausible assumption that $R_{0a}=R_{0a^c}$ (in any case
these quantities have
the same sign, see below)
we can
sum both sides over $a$. Then we get
\ba
\sum_{a,\alpha} R_{(m,\alpha)a}R_{(m,\alpha)a} &=&
\sum_{\ell,\alpha,\gamma} X_{(m,\alpha)(m,\alpha);(\ell,\gamma)}
\sum_a R_{(\ell,\gamma)a}R_{0a^c}  \nonumber \\
&=& \sum_{\alpha} X_{(m,\alpha)(m,\alpha);0}
 \label{eq:interma}
\ea

If we set $n=\beta=0$ in the classifying algebra (\ref{eq:classif}) , we obtain
\be R_{(m,\alpha)a}R_{0a}=\sum_{\ell,\gamma} X_{(m,\alpha)0;(\ell,\gamma)}
R_{(\ell,\gamma)a}R_{0a} \label{eq:classalgTWO} \ee
Now note that from the expression for the annulus amplitude we may
derive
$$ R_{0a}R_{0b}=S_{00} \sum_{j} A_{jab} S_{j0} $$
Therefore (since in unitary CFT's $S_{j0} > 0$)
$R_{0a}R_{0a^c} > S_{00} A_{0aa^c} S_{00}=(S_{00})^2 > 0$. Hence all
$R_{0a}$ are non-vanishing (and have the same sign).
So we can divide both sides of (\ref{eq:classalgTWO}) by $R_{0a}$
and find
$$ R_{(m,\alpha)a}=\sum_{\ell,\gamma} X_{(m,\alpha)0;(\ell,\gamma)}
R_{(\ell,\gamma)a}$$
This clearly implies
$$ X_{(m,\alpha)0;(\ell,\gamma)}=\delta_{m\ell}\delta_{\alpha\gamma} $$
We can also solve for all $X$'s in terms of $R$. The result is
$$  X_{(m,\alpha)(n,\beta);(\ell,\gamma)}
=\sum_a {R_{(m,\alpha)a} R_{(n,\beta)a} R_{(\ell,\gamma)a^c}\over R_{0a}} $$
Note that if all the boundaries are self-conjugate, this quantity is
symmetric in the three labels. Then
$$ X_{(m,\alpha)(m,\alpha);0}=X_{(m,\alpha)0;(m,\alpha)}=1$$
so that the sum over $\alpha$ in (\ref{eq:interma}) just gives $Z_{mm^c}$. On  
the other hand, in that
case $\sum_{\ell} Y^{\ell}_{~00}K_{\ell}$ must take its maximal value, since
with self-conjugate boundaries $\sum_a (M^0_a)^2$ equals the number of boundaries,
and hence the number of Ishibashi's. Hence we have
$$ \sum_i Z_{ii^c} = \sum_a (M^0_a)^2=\sum_{i} Y^{i}_{~00}K_{i} \leq \sum_i
Z_{ii^c}$$
Since the inequality must saturate, and since it holds for each $i$
separately, we clearly find
\be Y^{m}_{~~00}K_{m} =  Z_{mm^c}  = \sum_a \sum_{\alpha}
R_{(m,\alpha)a}R_{(m,\alpha)a} \label{eq:OrTrace} \ee

This estabishes the conjecture for real boundaries.
Note that in that case $\sum_a A^i_{~aa}=\sum_a A^{i~~a}_{~a}$, so that
the left hand sides of (\ref{eq:TraceOne}) and (\ref{eq:traceconj}) are identical. 
Nevertheless the second trace identity (\ref{eq:traceconj}) contains non-trivial
information, since it constrains (and in most cases fixes) the Klein
bottle coefficients $K^i$.

A further generalization
can also be proved, namely when boundary conjugation is non-trivial,
but is linked to charge conjugation in the bulk theory as
$R_{(\ell,\gamma)a^c}=R_{(\ell^c,\gamma)a}$. This
is true for instance in the Cardy case (\ie\ $R_{ma}=S_{ma}$),
even in complex CFT's.  Then we can derive
(\ref{eq:OrTrace}) for all real labels $m$. For complex $m$ on the one hand
$Y^{m}_{~00}=0$, whereas on the other hand $X_{(m,\alpha)(m,\alpha);0}=0$
vanishes because the classifying algebra coefficients vanish whenever
the corresponding fusion coefficients $N_{mm}^{~~0}$ vanish.
This gives an easy explanation for the fact that the only Klein bottle
choice consistent with the Cardy case is $K^i=Y^{i}_{~00}$.

What remains to be proved is the ``non-saturated" case, where some Klein
bottle coefficients are not equal to $Y^{i}_{~00}$. We were unable to
extend the foregoing derivation to such cases, but we did verify that
the conjecture holds for the class discussed in ~\cite{FOE}. In this paper
boundary and crosscap coeficients were presented for all simple current
modular invariants (multiplied by charge conjugation)
and (presumably) all consistent Klein bottle choices
for each invariant. Obviously this includes all non-trivial Klein bottle
choices for the charge conjugation invariant. A rather lengthy calculation,
which we will not present here, shows that indeed (\ref{eq:traceconj}) holds.
Another non-trivial test are
the results of ~\cite{NUNO} for $c=1$ orbifolds. In this case the Klein
bottle amplitude is non-standard, but (\ref{eq:traceconj}) nevertheless holds.

It appears that an essential ingredient in boundary CFT (by which we mean
conformal field theory on surfaces with boundaries {\it and} crosscaps)
is still missing. We clearly need a deeper understanding of the
completeness condition; furthermore a derivation of the trace formula
(\ref{eq:traceconj}) -- if indeed correct -- seems to require some additional  
insight.
It appears that the boundary and crosscap data fit tightly together, and
that one may be missing an important piece of the puzzle by focussing
only on boundary data, as is the case in most of the literature.
We hope that the trace formulas and polynomial equations we have
derived or conjectured provide a clue towards an underlying
structure. In any case, they are already useful for extending the
list of explicit solutions to exceptional cases. Needless to say,
we encourage explicit checks of our conjecture (\ref{eq:traceconj}), and would  
very much
like to hear about confirmations or counter examples.

\acknowledgments

A.N. Schellekens wishes to thank the theory group of Tor Vergata
for hospitality and the INFN for financial support.
Ya.S. wishes to thank NIKHEF for hospitality and financial support.
Some of our results were already presented
at the 1998 DESY workshop by the second
author, based on joint, unpublished work with A. Sagnotti \cite{Stanev}. The new  
results added in the
present paper include the extension from automorphisms to arbitrary modular
invariants and the orientation sensitive trace formulas.
We are especially
grateful to A. Sagnotti for numerous discussions. We would also like to thank
P.Bantay, L.R.Huiszoon, G.Pradisi and C.Schweigert for discussions
and comments.
The research of Ya.S. was supported in part
by the EEC contracts HPRN-CT-2000-00122
and HPRN-CT-2000-00148 and by the INTAS contract 99-1-590.


\begin{thebibliography}{999}

\bibitem{Cardy}
J. Cardy, Nucl. Phys. B324 (1989) 581.

\bibitem{Rome}
M. Bianchi and A.
Sagnotti, Phys. Lett. B247 (1990) 517; \\
D. Fioravanti, G. Pradisi and
A.Sagnotti, Phys. Lett. B321 (1994) 349;  \\
G. Pradisi, A. Sagnotti and Ya.S.
Stanev, Phys. Lett. B354 (1995) 279, Phys. Lett. B356 (1995)
 230.

\bibitem{SB}
J.~Fuchs and C. Schweigert, Phys. Lett.
B447 (1999) 266, Nucl. Phys. B558 (1999) 419; \\
A. Recknagel and V. Schomerus, Nucl. Phys. B531 (1998) 185.

\bibitem{FOE}
J. Fuchs, L.R. Huiszoon, A.N. Schellekens,
 C. Schweigert and J. Walcher, Phys. Lett. B495 (2000) 427.

\bibitem{Pmat}
G. Pradisi and A. Sagnotti, Phys. Lett.
B216 (1989) 59.

\bibitem{CMPLT}
G. Pradisi, A. Sagnotti and Ya.S. Stanev, Phys. Lett.
B381 (1996) 97.

\bibitem{DiFZ}
P. Di Francesco and J.-B Zuber, Nucl. Phys. B338 (1990)
602.

\bibitem{Zuber}
R. Behrend, P. Pearce, V. Petkova and J-B. Zuber,
Nucl. Phys. B570 (2000) 525, Nucl. Phys. B579 (2000) 707.

\bibitem{Gann}
T. Gannon, hep-th/0106105.

\bibitem{ClassAlg}
J.~Fuchs and C. Schweigert, Phys. Lett. B414 (1997) 251.

\bibitem{Ytens}
T. Gannon, Phys. Lett. B473 (2000) 80; see also \\
G. Pradisi, A. Sagnotti and Ya.S.
Stanev, Phys. Lett. B354 (1995) 279; \\
L.~Borisov, M.B.~Halpern and
C.~Schweigert, Int. J. Mod. Phys. A13 (1998) 125;
\\ L.R. Huiszoon, A.N. Schellekens and N.~Sousa,
Phys. Lett. B470 (1999) 95;
\\
P. Bantay, Phys. Lett. B419 (1998) 175 and
hep-th/9910079.

\bibitem{Kbtl}
L.R. Huiszoon, A.N. Schellekens and N.~Sousa,
Phys. Lett. B470 (1999) 95.

\bibitem{Bantay}
P. Bantay, Phys. Lett. B394 (1997) 87.

\bibitem{NUNO}
A.N. Schellekens and N.~Sousa, Int. J. Mod. Phys. A16 (2001) 3659.

\bibitem{Stanev}
Ya.S. Stanev, {\it Open descendants of Gepner models
in D=6}, talk given at workshop `Conformal Field Theory of D-branes' Sept. 7-12  
1998 at DESY, Hamburg, Germany. Transparancies on
http://www.desy.de/~jfuchs/CftD.html; \\
A. Sagnotti and Ya. S. Stanev, unpublished.


\end{thebibliography}
\end{document}